# Entanglement of two qubits mediated by a localized surface plasmon

Khachatur V. Nerkararyan[1] and Sergey I. Bozhevolnyi[2,*]

[1]*Department of Physics, Yerevan State University, 375049 Yerevan, Armenia*

[2]*Department of Technology and Innovation, University of Southern Denmark, Niels Bohrs Allé 1, DK-5230 Odense M, Denmark*

**Abstract**

We investigate relaxation dynamics in the system of two identical quantum dipole emitters (QDEs) located near a metal nanoparticle (MNP) exhibiting a dipolar localized surface plasmon (LSP) resonance at the frequency of the QDE radiative transition. Considering one QDE to be brought into an optically active excited state and weakly coupled to the resonant LSP, we show that a stable superposition state of two QDEs is formed during the transition time, which is much shorter than the QDE spontaneous decay time and determined by the efficiency of resonant interaction between the QDEs and induced LSP. It is elucidated that the superposition state is established as a result of redistribution of the energy of the initially excited QDE so that the corresponding steady-state QDE fields induced at the MNP site *cancel* each other. The degree of steady-state entanglement characterized by the concurrence is found dependent only on the ratio of distances between the QDEs and the MNP, reaching its maximum value of ~ 0.65, when the separation between the MNP and the initially excited QDE is larger by ~ 20% than the distance from the other QDE to the MNP.

# I. INTRODUCTION

Entanglement is a direct consequence of the superposition principle in quantum mechanics that implies, among other things, that a composite system can feature mixed states that cannot be factorized in products of states of its components. A two-qubit system is the simplest of composite systems, which displays many of the fundamentals features of quantum mechanics such as superposition and entanglement. Although many aspects of these features have by now been well studied,[1-6] the problem of significantly reducing the transition time, during which an entangled state is established, has been paid relatively little attention to. Meanwhile, the possibility of influencing and speeding up the transition process might become playing a crucial role in the future development of quantum information processing. It should be noted that the entangled state formation can be speeded up due to the Purcell effect as well as due to dealing with superradiant triplet states.[6] We believe that one of the possible approaches to reducing the characteristic time of formation of a superposition state is to make use of resonant elements with ultrafast response, even if the latter is achieved at the cost of strong dissipation.

The resonant coupling between quantum dipole emitters (QDEs), such as molecules or quantum dots, and a localized surface plasmon (LSP) of a metal nanoparticle (MNP) at optical frequencies allows control over the flow of electromagnetic energy and lies at the core of an explosively growing field of quantum plasmonics.[7] Recent advances in nano-optics, especially experiments with single molecules interacting with well-defined metal nanostructures,[8-10] often referred to as nanoantennas, serve as a strong impetus for further developments in this direction.[11,12] The most often discussed effect of QDE–MNP interaction is concerned with the modification (enhancement or quenching) of fluorescence yield determined by the balance between radiative and nonradiative decay rates, both enhanced near MNPs.[9,10,13-15] It is also expected that the QDE–MNP interaction can even enter the regime of strong coupling, where excitation energy is coherently transferred between QDE and MNP in the form of Rabi oscillations.[16]

The qubit-qubit coupling induced by propagating surface plasmon-polariton modes supported by one-dimensional waveguides has recently been theoretically considered.[6,17] In this paper, we demonstrate that two QDEs, which are resonantly coupled to the MNP supported LSP with only one QDE being initially excited, can form a stable coherent superposition state significantly faster than spontaneous emission of an isolated QDE. In this superposition state, non-zero dipole moments of the QDEs are *out of phase* with respect to each other, so that the total electric field acting on the MNP vanishes, eliminating thereby energy dissipation (i.e., field absorption in the MNP) and extending significantly the lifetime of the entangled state.

The considered system consisting of two identical QDEs and an MNP is characterized by the following important features:

1. The LSP decay rate is typically much larger than the QDE spontaneous emission (decay) rate, with the difference amounting up to five orders of magnitude. Therefore, even for relatively strong QDE-MNP interactions, the relaxation of the QDE-MNP-QDE system is much slower than the LSP decay, a feature that allows one to disregard the LSP dynamics and consider the MNP response as being *instantaneous*. At the same time, the QDE-MNP-QDE coupling is considered to be sufficiently strong so that the corresponding relaxation processes occur much faster than the isolated QDE relaxation, allowing us to disregard the QDE spontaneous emission.

2. The magnitude of a dipole moment associated with the radiative QDE transition is one order of magnitude smaller than that of an LSP dipole moment induced by the QDE, a feature that allows one to consider the MNP acting as an *antenna* of the QDE-MNP-QDE system.

3. Resonant excitation of the LSP is realized as a result of free electron oscillations in the MNP that, for nm-size MNPs, can be regarded as classical current oscillations, since a large number of free electrons (~100 nm$^{-3}$) is involved and their energy spectrum can be considered continuous. This classical oscillating current can then be represented by a quantum *coherent* state[18] of the LSP. Note that the coherent LSP state is fundamentally different from (often considered) LSP states with a definite number of quantized plasmons.[7]

The first two of the above properties create the conditions for the rapid formation of a superposition state and the efficient radiation with relatively small ohmic losses. The third feature of the QDE-MNP-QDE system determines the character of a quantized LSP field. From the basic principles of quantum optics, it is known that quantized fields created by classical currents are described by a wave function of the coherent state.[18] These quantized fields are largely equivalent to classical fields, allowing one to employ the semiclassical approximation. In particular, in the study of the relaxation dynamics of in the resonantly coupled QED-MNP system, it was found that the results obtained using both quantum[19] and semiclassical[20] approaches are identical when representing the LSP oscillating current by the coherent state in the quantum approach.[19] The same equivalence can also be established for the considered configuration (see Appendix), while the approach considered hereafter is essentially based on the semiclassical representation.

## II. THEORETICAL FRAMEWORK

The QDE-MNP-QDE system under consideration is schematically presented in Fig. 1, and consists of two identical three-level QDEs[15,16] located at the axis of symmetry on both sides of a spherical MNP. It is assumed that an external pump laser brings the first QDE, which we consider to be situated to the right from the MNP at distance $R_{10}$ in Fig. 1(a), from the ground state 0 into the excited state 2, where it decays nonradiatively into the optically active state 1 with energy $E_1$, while the second QDE situated to the right from the MNP at distance $R_{01}$ is, at this initial moment of time, in the ground state with energy $E_0$. It is further assumed that the spherical MNP exhibits a dipolar LSP resonance at the frequency $\omega_0$ of the radiative (dipole-allowed) transition $1 \to 0$ [Fig. 1(b)]. This allows us to separate the excitation dynamics, which is not influenced by the presence of the MNP, from the relaxation dynamics of the state 1 resulting in partial excitation of another QDE via the LSP field and formation of the coherent steady state, which is the main subject of this work. Physically, a very similar situation can be realized with two-level QDEs under two-photon (pulsed) excitation. Note that the shape of a MNP is not important in this context and can be chosen specifically in order to produce a dipolar resonance at a given frequency[16], e.g., to coincide with the QDE radiative transition frequency.

The wave function of the considered system can be represented the in general form:

$$\psi = b_{00}(t)\phi_{00}(\vec{r})e^{-\frac{i}{\hbar}(E_0+E_0)t} + b_{10}(t)\phi_{10}(\vec{r})e^{-\frac{i}{\hbar}(E_1+E_0)t} + b_{01}(t)\phi_{01}(\vec{r})e^{-\frac{i}{\hbar}(E_1+E_0)t} \quad , \quad (1)$$

where $\phi_{00}$, $\phi_{10}$, and $\phi_{01}$ are the wave functions of the QDE system, when respectively both QDEs are in the ground state, the first QDE is in the excited state while the second QDE is in the ground state, and when the first QDE is in the ground state while the second QDE is in the excited state; $b_{00}(t)$, $b_{10}(t)$ and $b_{01}(t)$ are the corresponding probability amplitudes. Here, it is taken into account that, since only one QDE is in the excited state at the initial time moment, the probability of both QDEs to be in the excited state is zero. The dipole moment of the first (second) QDE can then be written as follows:

$$\vec{D}_{10(01)} = b_{10(01)}b_{00}^*\vec{d}\exp[-i(\omega t - \varphi_{10(01)})] + c.c. \quad , \quad (2)$$

where

$$\vec{d}\exp(i\varphi_{10(01)}) = \int_V \phi_{10(01)} e\vec{r}\phi_{00}^* dV \quad , \quad \hbar\omega_0 = E_1 - E_0 \quad . \tag{3}$$

Hereafter we consider the optimum (from the viewpoint of efficient QDE-MNP coupling) orientation of the transition dipole moment, $\vec{d}$, along the symmetry axis connecting the MNP center with the QDEs [Fig. 1(a)], with $\varphi_{10}$ and $\varphi_{01}$ being initial (undetermined) phases. Let us further assume that the MNP center-to-QDE distances $R_{10}$ and $R_{01}$ are considerably larger than the MNP radius $r$ [Fig. 1(a)], with all dimensions being much smaller than the wavelength $\lambda$ of light, i.e., that $\lambda \gg \max[R_{10}, R_{01}]$ and $r \ll \min[R_{10}, R_{01}]$. In this electrostatic approximation, the MNP can be considered as being subjected to the homogenous electric field $\vec{E}_{sph}$ created by the oscillating QDE dipoles:

$$\vec{E}_{sph} = \frac{2}{4\pi\varepsilon_0\varepsilon_2}\left[\frac{\vec{D}_{10}}{R_{10}^3} + \frac{\vec{D}_{01}}{R_{01}^3}\right] = \vec{E}_0 e^{-i\omega t} + \text{c.c.} \quad . \tag{4}$$

Here, c.c. stands for complex conjugate, $\varepsilon_0$ and $\varepsilon_2$ are the relative permittivities of vacuum and the dielectric environment. In general, the MNP response should be determined by considering the corresponding dynamics influenced by the external field and the LSP relaxation [see Eq. (A14)]. However, as elucidated in the introduction (and noted in our previous work[19,20]), the MNP response can be considered instantaneous due to extremely fast relaxation of LSP excitation [see also Eq. (A17)] and employ the electrostatic approximation for its description. The resonant LSP induced in the MNP by the QDE induced field creates in its turn the electric fields at the QDE sites that can be written in the following form:

$$\vec{E}_{10(01)} = \frac{2(\varepsilon_1 - \varepsilon_2)r^3}{(\varepsilon_1 + 2\varepsilon_2)R_{10(01)}^3}\vec{E}_0 e^{-i\omega t} + \text{c.c.} \quad , \tag{5}$$

where $\varepsilon_1 = \varepsilon_{1r} + i\varepsilon_{1i}$ is the MNP relative permittivity.

Using the time-dependent Schrödinger equation for two-level systems in the driving field given by Eq. (5) and carrying out standard manipulations within the rotating wave approximation, one obtains the following system of coupled equations for the probability amplitudes ($\dot{b} \equiv db/dt$):

$$\dot{b}_{10} = \frac{i(\varepsilon_1 - \varepsilon_2)|\vec{d}|^2 r^3}{\pi \hbar \varepsilon_0 \varepsilon_2 (\varepsilon_1 + 2\varepsilon_2) R_{10}^3} \left[ \frac{|b_{00}|^2 b_{10}}{R_{10}^3} + \frac{|b_{00}|^2 b_{01}}{R_{01}^3} e^{i(\varphi_{01} - \varphi_{10})} \right] , \qquad (6)$$

$$\dot{b}_{01} = \frac{i(\varepsilon_1 - \varepsilon_2)|\vec{d}|^2 r^3}{\pi \hbar \varepsilon_0 \varepsilon_2 (\varepsilon_1 + 2\varepsilon_2) R_{01}^3} \left[ \frac{|b_{00}|^2 b_{01}}{R_{01}^3} + \frac{|b_{00}|^2 b_{10}}{R_{10}^3} e^{-i(\varphi_{01} - \varphi_{10})} \right] , \qquad (7)$$

$$\dot{b}_{00} = \frac{i(\varepsilon_1^* - \varepsilon_2)|\vec{d}|^2 r^3}{\pi \hbar \varepsilon_0 \varepsilon_2 (\varepsilon_1^* + 2\varepsilon_2)} \left[ \frac{b_{10}}{R_{10}^3} e^{i\varphi_{10}} + \frac{b_{01}}{R_{01}^3} e^{i\varphi_{01}} \right] \left[ \frac{b_{10}^*}{R_{10}^3} e^{-i\varphi_{10}} + \frac{b_{01}^*}{R_{01}^3} e^{-i\varphi_{01}} \right] b_{00} . \qquad (8)$$

In obtaining the above relations, we assumed that temporal variations of $b_{00}$, $b_{10}$ and $b_{01}$ during the LSP lifetime are insignificant, i.e. that the QDE-MNP-QDE dynamics is very slow in comparison with the LSP damping, an assumption that is consistent with the weak-coupling regime as elucidated in the introduction.

The obtained equations can be further simplified and made amenable to analytical treatment by considering the resonance configuration and relatively low LSP damping, i.e., with the following conditions being satisfied: $|\varepsilon_{1r} + 2\varepsilon_2| << \varepsilon_{1i}$ and $3\varepsilon_2 >> \varepsilon_{1i}$. In this case, the coupled equations become reduced to:

$$\dot{b}_{10} = q[\beta_{10} b_{10} + \beta_{01} b_{01}] \beta_{10} b_{00}^2 , \qquad (9)$$

$$\dot{b}_{01} = -q[\beta_{01} b_{01} + \beta_{10} b_{10}] \beta_{01} b_{00}^2 , \qquad (10)$$

$$\dot{b}_{00} = q[\beta_{10} b_{10} + \beta_{01} b_{01}]^2 b_{00} , \qquad (11)$$

with the initial phases being incorporated into the corresponding probability amplitudes: $b_{10(01)} \exp(i\varphi_{10(01)}) \Rightarrow b_{10(01)}$, and the following notation being introduced:

$$q = \frac{3r^3 |\vec{d}|^2}{\pi \varepsilon_0 \varepsilon_{1i} \hbar} \frac{(R_{10}^6 + R_{01}^6)}{R_{10}^6 R_{01}^6} \quad \text{and} \quad \beta_{10(01)} = \frac{R_{01(10)}^3}{\sqrt{R_{10}^6 + R_{01}^6}} . \qquad (12)$$

Let us consider the initial moment $\tau$ of time being characterized with the following conditions: $b_{10}(\tau) \approx 1$, $b_{00}(\tau) = \chi << 1$, and $b_{01}(\tau) = 0$. In other words, we consider the coupling processes in the QDE-MNP-QDE system, which are described by the above equations, to commence when the first (initially excited) QDE is partially relaxed into the ground state. This starting process can occur due to other inducements always found in the open system, for example, due to the free-space spontaneous emission *without* interacting with the MNP because the MNP dipole moment, which can only be induced via the first QDE relaxation, is negligibly small. Under the specified initial conditions, the above system of equations has the following solutions:

$$b_{10}(t) = \frac{\beta_{10}^3}{\sqrt{\beta_{10}^2 + \chi^2 e^{2\mu(t-\tau)}}} + \beta_{01}^2 \quad , \tag{13}$$

$$b_{01}(t) = \frac{\beta_{10}^2 \beta_{01}}{\sqrt{\beta_{10}^2 + \chi^2 e^{2\mu(t-\tau)}}} - \beta_{10}\beta_{01} \quad , \tag{14}$$

$$b_{00}(t) = \frac{\beta_{10} \chi e^{\mu(t-\tau)}}{\sqrt{\beta_{10}^2 + \chi^2 e^{2\mu(t-\tau)}}} \quad , \tag{15}$$

with the characteristic temporal rate $\mu$ being as follows

$$\mu = \frac{3|\vec{d}|^2}{\pi \varepsilon_0 \varepsilon_{1i} \hbar} \frac{r^3}{R_{10}^6} \quad . \tag{16}$$

Note that the obtained solutions [Eqs. (13)-(15)] satisfy the condition: $|b_{10}|^2 + |b_{01}|^2 + |b_{00}|^2 = 1$.

The rate $\mu$ [Eq. (16)] of the considered process determines the characteristic relaxation time of the first QDE excited state. One of the most important assumptions made is related to the strength of the QDE-MNP coupling which should ensure considerably larger relaxation rates than that for the QDE in free space. Their ratio can be evaluated now with the help of Eq. (16) and the Weisskopf-Wigner result[21] as follows:

$$\beta \cong \frac{\mu}{\Gamma} = \frac{9}{\varepsilon_{1i}\sqrt{\varepsilon_2}} \left(\frac{\lambda_0}{2\pi R_{10}}\right)^3 \left(\frac{r}{R_{10}}\right)^3 \quad , \tag{17}$$

with $\lambda_0$ being the vacuum wavelength corresponding to the QDE transition frequency $\omega_0$. For a typical dielectric environment with $\varepsilon_2 = 2.25$ (e.g., glass or polymer), the resonance condition (i.e., $\varepsilon_{1r} = -4.5$) is met, for gold, at the wavelength of ~ 530 nm with $\varepsilon_{1i}^g \cong 2.35$ and, for silver, at ~ 400 nm with $\varepsilon_{1i}^{s} \cong 0.22$.[22] Considering an MNP with the radius of 5 nm and the first QDE distance to the MNP center being 15 nm (in order to be within the electrostatic dipole description), one obtains the ratio $\beta \approx 17$ for gold and $\beta \approx 77$ for silver, justifying thereby the aforementioned assumption: $\mu \gg \Gamma$. It is interesting that the effect is already pronounced at relatively large (~ 10 nm) distances between QDEs and the MNP surface, which are in the range of distances explored in the recent experiments with 10-nm-size gold nanoparticles.[14] It is also transparent that even larger ratios can be achieved by exploiting the LSP shape dependence[16] and red-shifting the MNP resonance towards smaller metal absorption.[22]

The dynamics of the first QDE relaxation and its coupling to the second QDE via the LSP excitation, which is described by Eqs. (14)-(16), starts off when nonzero population of the first QDE ground state is reached due to other (relatively slow) relaxation processes with an exponential decay, so that $b_{10}^{r}(t) = \exp(-\Gamma t)$. Applying the continuity condition at the transition between these two processes to both functions, $b_{10}^{r}(t)$ and $b_{10}(t)$, and their derivatives, one can determine the characteristic time $\tau = 1/2\mu$, after which the role of the investigated process will be dominant. Note that this initial time does not depend on the QDE relaxation rate $\Gamma$ in free space. This procedure allows studying the behavior of the QDE-MNP-QDE system during the whole process.

### III. RESULTS AND DISCUSSION

The most remarkable feature of the considered process is the formation of a superposition state. Indeed, it is seen [Eqs. (13)-(15)] that the probability amplitudes evolve towards stationary values: $b_{10} \to \beta_{01}^2$, $b_{01} \to -\beta_{10}\beta_{01}$ and $b_{00} \to \beta_{10}$. The occurrence of the superposition state with non-zero QDE transition dipole moments in the presence of a strongly dissipating LSP seems *counterintuitive*. The explanation is nevertheless rather straightforward: as follows from the expression for the field $\vec{E}_{sph}$ created by the oscillating QDE dipoles at the site of the MNP [Eq. (4)], the total QDE field acting on the MNP *vanishes* at the end of the transition period: $\vec{E}_{sph} \to 0$ due to the QDE dipole fields (at the MNP) becoming exactly equal and out of phase. It is a direct consequence of the LSP resonance occurring *exactly* at the QDE radiative transition frequency that the MNP fields acting back on the QDEs [Eq. (5)] are $\pi/2$ phase-shifted with respect to the field $\vec{E}_{sph}$ acting on the MNP [Eq. 4)]. This phase shift results, in turn, in the remarkable feature of the QDE-MNP-QDE system evolution towards the superposition state with the QDE dipole moments being out of phase [Eq. (2)] and, consequently, their total field $\vec{E}_{sph}$ at the site of the MNP [Eq. (4)] being extinguished, i.e., the LSP being quenched.

The evolution of the perfectly symmetrical QDE-MNP-QDE system (i.e., with $R_{10} = R_{01}$) towards the stable superposition state (Fig. 2) indicates that, for a reasonably large speed-up factor $\beta = 100$, the steady state is achieved during the time period of $\approx 0.05/\Gamma$, i.e., $\approx 20$ times faster than the QDE spontaneous emission in free space. Approximately the same time is needed to completely extinguish the total QDE field acting on the MNP (Fig. 3) and, thereby, quench the LSP. It is therefore realistic (see above estimations of $\beta = \mu/\Gamma$) to create a stable superposition state existing sufficiently long time, i.e., until the process of spontaneous emission commences.

The steady-state probability amplitudes depend strongly on the asymmetry in the QDE positions with respect to the MNP, i.e., on the ratio $R = R_{10}/R_{01}$ (Fig. 4), as can also be perceived from Eqs. (12)-(15). The electromagnetic interactions in the QDE-MNP-QDE system are short-range (or near-field) electrostatic interactions, and even relatively small differences in the QDE-MNP distances result in large variations of the probabilities of different system states (Fig. 4). For substantial differences, the system steady state becomes rapidly characterized by either both QDEs being in the ground state (for $R < 0.5$) or the originally excited QDE preserving its excited state (for $R > 2$). It should be understood that all various steady states considered here would stay intact only within a fraction of the QDE lifetime in free space. After that, the process of QDE spontaneous emission commences and, therefore, can no longer be ignored.

In the considered configuration, the two QDEs are characterized by the coherent superposition of their states [Eq. (1)] or, in other words, entangled. The degree of entanglement can conveniently be characterized by the concurrence, which in our case is given by $C = 2|b_{10}b_{01}|$.[6,23] During the formation of the superposition state, the concurrence increases monotonically reaching its maximum at the steady state that depends on the ratios $R = R_{10}/R_{01}$ between the QDE-MNP distances (Fig. 5). Using the above formulae [Eqs. (12)-(15)], one can obtain the following simple relation for the concurrence at the steady state condition:

$$C \to 2\beta_{10}\beta_{01}^3 = \frac{2R^9}{\left(1+R^6\right)^2} \quad . \tag{18}$$

It is can be shown, using Eq. (18), that the steady-state concurrence attains the maximum value of $C_{max} = 3\sqrt{3}/8 \cong 0.65$ for the ratio $R_{opt} = \sqrt[6]{3} \cong 1.2$ (Fig. 6). Even though the steady-state concurrence for the perfectly symmetrical QDE-MNP-QDE configuration is not maximal, $C(R=1) = 0.5$, this configuration is very interesting because its steady state is completely antisymmetric: $b_{10} = -b_{01} = 0.5$, with the QDE dipole moments being equal and oppositely oriented [Eq. (2)]. Finally, we would like to point out that, since the losses by absorption (ohmic losses) are inevitable in any plasmonic system, the concurrence is expected to be limited: $C < 1$.

The levels of concurrence predicted for our configuration are comparable with those calculated for the qubit entanglement mediated by channel plasmons,[6] albeit the entanglement mechanism as well as its decay is rather different. While everything that we describe occurs via near-field (electrostatic) QDE interactions (mediated by the LSP excitation), the entanglement via channel plasmons in realistic configurations is associated with the channel plasmon absorption (during propagation) resulting in the decay of entanglement starting at the *very beginning* of the process, limiting thereby the distance between two qubits that can be used to ensure reasonable levels of entanglement.[6] At the same time, it is worth noting that the *actual loss* characteristics, such as the imaginary part of the susceptibility, do not appear explicitly in the expression for the steady-state concurrence [Eq. (18)]. This remarkable feature is, in our opinion, related to a somewhat similar attribute of the resonantly coupled QDE-MNP system that we considered recently, namely, to the QDE-MNP absorption efficiency (probability of excitation energy dissipation by the MNP absorption) being ~25% irrespectively of the MNP dielectric properties.[19]

## IV. CONCLUSIONS

We have reported a semiclassical consideration of relaxation dynamics in the system of two identical QDEs located near a MNP exhibiting a dipolar LSP resonance at the frequency of the QDE radiative transition. Considering one QDE to be brought into an optically active excited state (for example, by a short pump pulse exciting a higher level rapidly and nonradiatively decaying to the active state) and weakly coupled to the resonant LSP, we show that a stable superposition state of two QDEs is formed during the transition time, which is much shorter than the QDE spontaneous decay time and determined by the efficiency of resonant interaction between the QDEs and induced LSP. It is elucidated that the superposition state is established as a result of redistribution of the energy of the initially excited QDE so that the corresponding steady-state QDE fields induced at the MNP site *cancel* each other. The degree of steady-state entanglement characterized by the concurrence is found dependent only on the ratio of distances between the QDEs and the MNP, reaching its maximum value of ~ 0.65, when the separation between the MNP and the initially excited QDE is larger by ~ 20% than the distance from the other QDE to the MNP. The most intriguing feature of the considered process is that, despite the nonzero dissipation in the QDE-MNP-QDE system (due to radiation absorption by the MNP), the steady state entanglement remains *completely unchanged* during the time that is much longer than the system characteristic time, until much slower process of spontaneous emission commences.

Concluding, we would also like to comment on the condition of matching the frequencies of the QDE radiative transition and the LSP resonance. This condition is critical only within the spectral width of the LSP resonance, which is rather large due to extremely fast LSP relaxation. Qualitative considerations suggest that a relatively small detuning would mainly result in a shift of energy levels, but this effect requires a special consideration. Finally, we should note that similar results can also be obtained by representing the LSP by a coherent state,[19] a task that is delegated to the Appendix. Overall, we believe that the reported results have far reaching implications within the very rapidly developing field of quantum plasmonics.

## ACKNOWLEDGMENTS

The authors gratefully acknowledge financial support for this work from the European Research Council, Grant No. 341054 (PLAQNAP), as well as partial support (KVN) from the Danish Council for Independent Research (Contract No. 09-072949, ANAP).

**APPENDIX: CONSIDERATION OF RELAXATION DYNAMICS USING THE DENSITY-MATRIX FORMALISM AND A COHERENT PLASMON STATE**

The Hamiltonian of the considered system (Fig. 1) can be represented as follows:

$$\hat{H} = \hat{H}_0 + \hat{H}_1 , \qquad (A1)$$

$$\hat{H}_0 = \frac{1}{2}\hbar\omega_0 \left(b_{e1}^+ b_{e1} + b_{e2}^+ b_{e2}\right) + \hbar\omega_p a^+ a , \qquad (A2)$$

$$\hat{H}_1 = \eta_1 \left(b_{e1}^+ b_g + b_g^+ b_{e1}\right)a + \eta_1^* \left(b_{e1}^+ b_g + b_g^+ b_{e1}\right)a^+ + \eta_2 \left(b_{e2}^+ b_g + b_g^+ b_{e2}\right)a + \eta_2^* \left(b_{e2}^+ b_g + b_g^+ b_{e2}\right)a^+ . \qquad (A3)$$

Here, $\omega_0$ is the frequency of the resonant QDE transition, $b_{e1}^+$ ($b_{e2}^+$) and $b_{e1}$ ($b_{e2}$) are the creation and annihilation operators of the excited state of the first (second) QED, and $b_g^+$ and $b_g$ are the creation and annihilation operators of the ground QDE states, $\omega_p$ is the frequency of the resonant LSP excitation, $a^+$ and $a$ are the creation and annihilation operators of the LSP, $\eta_1$ and $\eta_2$ are the coupling constants characterizing the interaction between the (first and second) QDEs:

$$\eta_{1(2)} = \frac{1}{4\pi\varepsilon_0\varepsilon_2}\frac{2\vec{d}\cdot\vec{d}_p}{R_{10(01)}^3} = \eta_0 \beta_{10(01)} , \qquad (A4)$$

where $\vec{d}_p$ is the dipole moment associated with the LSP transition:[24]

$$\left|\vec{d}_p\right|^2 = \frac{12\pi\hbar\varepsilon_0\varepsilon_2^2 r^3}{\partial\varepsilon_{1r}(\omega_p)/\partial\omega_p} , \quad \text{and} \quad \eta_0 = \frac{2\vec{d}\cdot\vec{d}_p}{4\pi\varepsilon_0\varepsilon_2}\frac{\sqrt{R_{10}^6 + R_{01}^6}}{R_{10}^3 R_{01}^3} . \qquad (A5)$$

By using the unitary transformation: $U_0 = \exp(-i\hat{H}_0 t/\hbar)$, we transform the system Hamiltonian into one that, within the rotation wave approximation and under the condition of strict resonance, $\omega_0 = \omega_p$, has the following form:

$$\hat{H}' = \eta_0 a\left(\beta_{10} b_{e1}^+ b_g + \beta_{01} b_{e2}^+ b_g\right) + \eta_0^* a^+ \left(\beta_{10} b_g^+ b_{e1} + \beta_{01} b_g^+ b_{e2}\right) . \qquad (A6)$$

Introducing new creation and annihilation operators corresponding to the singlet and triplet states formed by two coupled QDEs:

$$c_+^+ = \beta_{10} b_{e1}^+ + \beta_{01} b_{e2}^+ , \qquad c_-^+ = \beta_{10} b_{e1}^+ - \beta_{01} b_{e2}^+ , \qquad (A7)$$

$$c_+ = \beta_{10}b_{e1} + \beta_{01}b_{e2} \quad , \qquad c_- = \beta_{10}b_{e1} - \beta_{01}b_{e2} \quad , \tag{A8}$$

results in the modified system Hamiltonian:

$$\hat{H}_m = \eta_0 a c_+^+ b_g + \eta_0^* a^+ b_g^+ c_+ \quad . \tag{A9}$$

Considering free electron oscillations in the resonantly excited MNP as classical current oscillations (due to a very large number of electrons involved and the continuity of their energy spectrum), we further make use of the concept of coherent states in quantum optics[18] for the description of this classical current. We also assume that, due to an extremely large difference in the decay rates of an LSP and isolated QDE, it is possible and, indeed, highly probable that the relaxation of the QDE-MNP-QDE system is much slower than that of the LSP, but much faster than that of the isolated QDE. In such a situation, one can neglect the QDE relaxation due to its spontaneous emission and disregard the LSP dynamics, considering the MNP response as instantaneous. Under these conditions, the wave function of the full QDE-MNP system can be represented as follows:

$$|\psi\rangle = \left\{\left[\gamma_0(t) + \gamma_+(t)c_+^+ + \gamma_-(t)c_-^+\right]|g\rangle\right\} \exp\left(-\frac{1}{2}|\alpha(t)|^2\right) \sum_{n=0}^{\infty} \frac{\alpha^n(t)}{\sqrt{n!}}|n\rangle \quad . \tag{A10}$$

Here $\gamma_0(t)$, $\gamma_+(t)$, and $\gamma_-(t)$ are the probability amplitudes and $|g\rangle$ is the wave function of the system when both QDEs are in the ground state, $\alpha(t)$ is the eigenvalue of the operator $a$ and $|n\rangle$ is the LSP wave function corresponding to the energy eigenvalue $n\hbar\omega_p$. Eqs. (A9) and (A10) allow us to finally obtain relations for the probability density matrix elements of the QDE transitions and for the eigenvalue $\alpha(t)$ of the LSP operator:

$$\frac{d|\gamma_+(t)|^2}{dt} = \frac{d\rho_{++}}{dt} = \frac{i}{\hbar}\left(\eta_0 \alpha^* \rho_{0+} - \eta_0^* \alpha \rho_{+0}\right) \quad , \tag{A11}$$

$$\frac{d\gamma_+^*\gamma_0}{dt} = \frac{d\rho_{+0}}{dt} = \frac{d\rho_{0+}^*}{dt} = -\frac{i}{\hbar}\eta_0 \alpha^*\left(\rho_{++} - \rho_{00}\right) \quad , \tag{A12}$$

$$\frac{d|\gamma_-(t)|^2}{dt} = \frac{d\rho_{--}}{dt} = 0 \quad , \qquad \rho_{++} + \rho_{--} + \rho_{00} = 1 \quad , \tag{A13}$$

$$\frac{d\alpha(t)}{dt} = -\frac{i}{\hbar}\eta_0 \rho_{0+} - \alpha \Gamma_{LSP} \quad , \tag{A14}$$

where the LSP relaxation rate $\Gamma_{LSP}$ is introduced:[16,24]

$$\Gamma_{LSP} = \frac{\varepsilon_{1i}(\omega_p)}{\partial \varepsilon_{1r}(\omega_p)/\partial \omega_p} \quad , \tag{A15}$$

and

$$\rho_{++} = \left\langle c_+^+ c_+ \right\rangle, \quad \rho_{--} = \left\langle c_-^+ c_- \right\rangle, \quad \rho_{00} = \left\langle b_g^+ b_g \right\rangle, \quad \rho_{+-} = \left\langle c_+^+ b_g \right\rangle = \rho_{-+}^* \quad . \tag{A16}$$

Here, we operate under the assumption of instantaneous MNP response: $d\alpha/dt \ll \alpha \Gamma_{LSP}$, hence:

$$\alpha \approx -\frac{i}{\hbar \Gamma_{LSP}}\eta_0 \rho_{0+} \quad . \tag{A17}$$

Finally, one can work out the following solution of the system of Eqs. (A11-A14):

$$\rho_{++}(t) = \frac{\beta_{10}^2}{1 + e^{2[\mu(t-\tau)-\vartheta_0]}}, \quad \rho_{--}(t) = \beta_{01}^2, \quad \rho_{00}(t) = \frac{\beta_{10}^2 e^{2[\mu(t-\tau)-\vartheta_0]}}{1 + e^{2[\mu(t-\tau)-\vartheta_0]}}, \tag{A18}$$

$$\rho_{+0}(t) = \rho_{0+}(t) = \frac{\beta_{10}^2}{2\cosh[\mu(t-\tau)-\vartheta_0]}, \quad \mu = \frac{3|\vec{d}|^2}{\pi \varepsilon_0 \varepsilon_{1i} \hbar}\frac{r^3}{R_{10}^6}, \quad \vartheta_0 = \frac{1}{2}\ln\frac{\rho_{++}^\tau}{\rho_{00}^\tau} \quad , \tag{A19}$$

where $\rho_{++}^\tau$ and $\rho_{00}^\tau$ are the matrix elements at the initial moment of time $t = \tau$. Finally, using Eqs. (A7-A10) one obtains:

$$\gamma_+ = \beta_{10}b_{10} + \beta_{01}b_{01}, \quad \gamma_- = \beta_{01}b_{10} - \beta_{10}b_{01} \quad . \tag{A20}$$

It is quite straightforward to detect the complete equivalence between the solutions obtained in the main text [Eqs. (13-15)] and Eqs. (A18-A19), noticing that $\vartheta_0 = \ln(\beta_{10}/\chi)$.

**Figure captions**

FIG. 1. (Color online) Schematic of a system with two QDEs placed near an MNP, indicating (a) system parameters and (b) relevant QDE energetic levels along with an oscillating current associated with the LSP excitation.

FIG. 2. (Color online) The magnitudes of probability amplitudes of the QDE excited states $b_{10}$ (solid line) and $b_{01}$ (dashed line) and of the ground state $b_{00}$ (dotted line) as a function of time normalized by the QDE relaxation rate $\Gamma$ in free space for $\beta = \mu/\Gamma = 100$ and $R_{10} = R_{01}$.

FIG. 3. (Color online) The electric field magnitude (in arbitrary units) created at the site of the MNP by the oscillating QDE dipoles as a function of normalized time (see caption to Fig. 2) for $\beta = 100, 20, 10$ (solid, dashed and dotted lines, respectively) and $R_{10} = R_{01}$.

FIG. 4. (Color online) The magnitudes of probability amplitudes of the QDE excited states $b_{10}$ (solid line) and $b_{01}$ (dashed line) and of the ground state $b_{00}$ (dotted line) as a function of the QDE-MNP distance ratio $R = R_{10}/R_{01}$ for the QDE-MNP-QDE system steady states, i.e. at the completion of the relaxation process.

FIG. 5. (Color online) The concurrence as a function of normalized time (see caption to Fig. 2) for different QDE-MNP distance ratios: $R = R_{10}/R_{01} = 1.2, 1, 2$ (solid, dashed and dotted lines, respectively).

FIG. 6. (Color online) The concurrence as a function of the QDE-MNP distance ratio $R = R_{10}/R_{01}$ at the completion of the relaxation process.

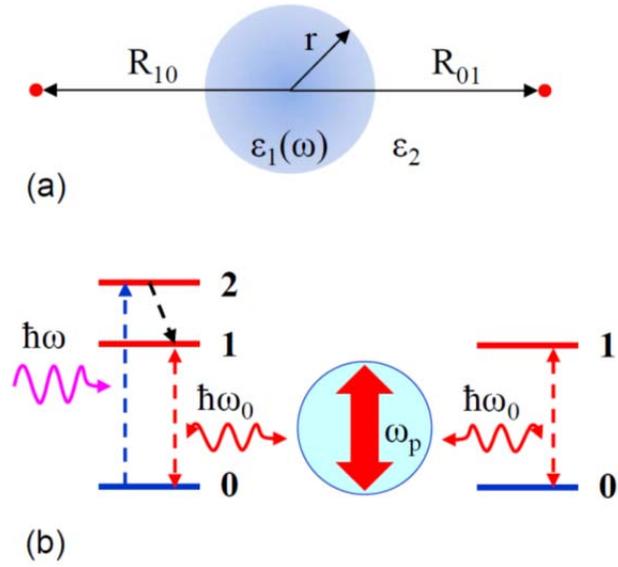

FIG. 1. (Color online) Schematic of a system with two QDEs placed near an MNP, indicating (a) system parameters and (b) QDE energetic levels along with an oscillating current associated with the LSP excitation.

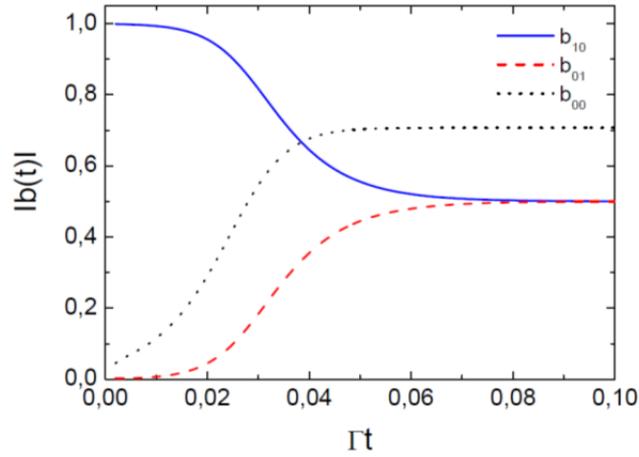

FIG. 2. (Color online) The magnitudes of probability amplitudes of the QDE excited states $b_{10}$ (solid line) and $b_{01}$ (dashed line) and of the ground state $b_{00}$ (dotted line) as a function of time normalized by the QDE relaxation rate $\Gamma$ in free space for $\beta = \mu/\Gamma = 100$ and $R_{10} = R_{01}$.

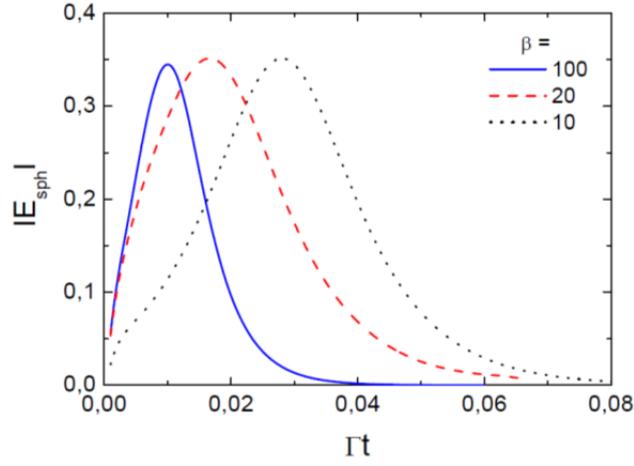

FIG. 3. (Color online) The electric field magnitude (in arbitrary units) created at the site of the MNP by the oscillating QDE dipoles as a function of normalized time (see caption to Fig. 2) for $\beta = 100, 20, 10$ (solid, dashed and dotted lines, respectively) and $R_{10} = R_{01}$.

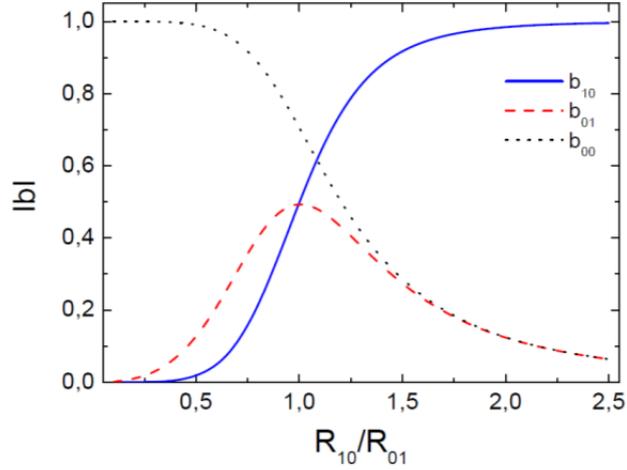

FIG. 4. (Color online) The magnitudes of probability amplitudes of the QDE excited states $b_{10}$ (solid line) and $b_{01}$ (dashed line) and of the ground state $b_{00}$ (dotted line) as a function of the QDE-MNP distance ratio $R = R_{10} / R_{01}$ for the QDE-MNP-QDE system steady states, i.e. at the completion of the relaxation process.

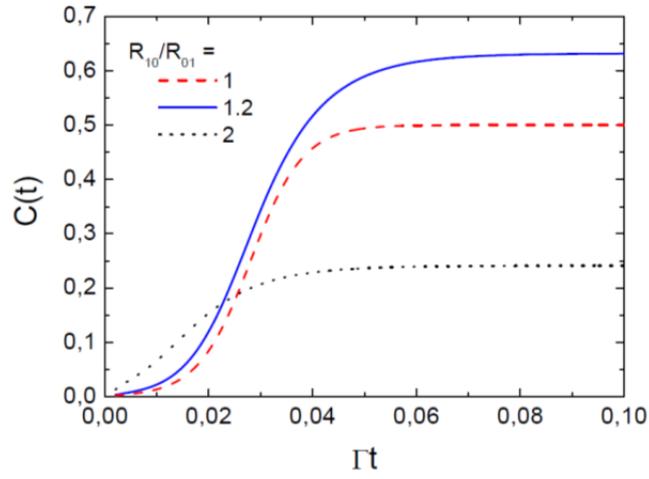

FIG. 5. (Color online) The concurrence as a function of normalized time (see caption to Fig. 2) for different QDE-MNP distance ratios: $R = R_{10}/R_{01} = 1.2, 1, 2$ (solid, dashed and dotted lines, respectively).

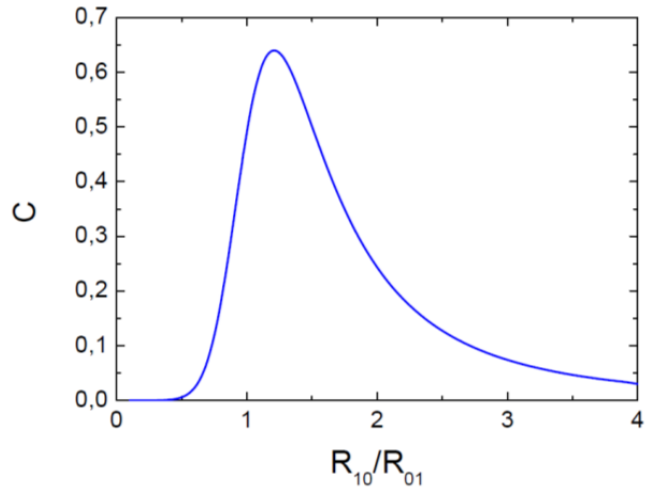

FIG. 6. (Color online) The concurrence as a function of the QDE-MNP distance ratio $R = R_{10}/R_{01}$ at the completion of the relaxation process.